%% file: main_cogsci_edited.tex
\documentclass[10pt,letterpaper]{article}

\usepackage{cogsci}

\cogscifinalcopy

\usepackage[style=apa,natbib=true,annotation=false]{biblatex}
\usepackage{float}
\usepackage{amsmath}
\usepackage{graphicx}

\title{Neural Fields as World Models}

\author[1,2]{\mbox{Joshua Nunley (joshnunl@iu.edu)}}
\affil[1]{Luddy School of Informatics, Computing, and Engineering, Indiana University, Bloomington}
\affil[2]{Cognitive Science Program, Indiana University, Bloomington}

\begin{document}

\maketitle

% Auto-generated results macros
\input{results_macros}

\begin{abstract}
Humans rehearse possible futures offline, as in mental practice and perhaps dreaming, suggesting that world models may support task learning away from the environment.
Standard machine learning world models compress visual input into latent vectors, discarding the spatial structure that characterizes sensory cortex.
We propose isomorphic world models: architectures that preserve sensory topology, so physics prediction becomes geometric propagation rather than abstract state transition.
We implement this idea with motor-gated neural fields, where activity evolves through local lateral connectivity and motor commands multiplicatively modulate specific channels.
Across three experiments, the same architecture learns ballistic prediction without ``teleporting,'' improves a catching policy offline by propagating task error through a frozen learned world model, and develops body-selective motor channels without body labels.
These results provide preliminary evidence that physical prediction, offline task learning, and body-linked representation share a common computational substrate: action-conditional prediction within a spatial map.

\textbf{Keywords:} neural fields; world models; intuitive physics; motor control; body schema
\end{abstract}

\input{sections/introduction_cogsci_edited}
\input{sections/methods_cogsci_edited}
\input{sections/results_cogsci_edited}
\input{sections/discussion_cogsci_edited}

\section{Acknowledgments}
This work used the Big Red 200 supercomputer at Indiana University, supported by Lilly Endowment, Inc., through its support for the Indiana University Pervasive Technology Institute. The author thanks Brandon Nunley for helpful discussions.
AI tools assisted with editing the manuscript and writing portions of the code; the author reviewed all AI-generated content, takes full responsibility for the final work, and validated the code through systematic testing and visualization.

\printbibliography

\end{document}

%% file: results_macros.tex
% Auto-generated by paper/compute_results.py
% Do not edit manually - regenerate with: python paper/compute_results.py --latex

% =============================================================================
% Ballistic Prediction Task
% =============================================================================
\newcommand{\ballisticNFPredLossMedian}{0.000933}
\newcommand{\ballisticNFPredLossMean}{0.000919}
\newcommand{\ballisticNFPredLossStd}{0.000152}
\newcommand{\ballisticNFPredLossIQRLow}{0.000895}
\newcommand{\ballisticNFPredLossIQRHigh}{0.000988}
\newcommand{\ballisticNFPredLossMedianSci}{9.33 \times 10^{-4}}
\newcommand{\ballisticVAEPredLossMedian}{0.003941}
\newcommand{\ballisticVAEPredLossMean}{0.003927}
\newcommand{\ballisticVAEPredLossStd}{0.001185}
\newcommand{\ballisticVAEPredLossIQRLow}{0.003447}
\newcommand{\ballisticVAEPredLossIQRHigh}{0.004528}
\newcommand{\ballisticVAEPredLossMedianSci}{3.94 \times 10^{-3}}
\newcommand{\ballisticTestU}{0}
\newcommand{\ballisticTestP}{0.0002}
\newcommand{\ballisticTestPFormatted}{p < 0.001}

% =============================================================================
% Puck Prediction Task
% =============================================================================
\newcommand{\puckNFPredLossMedian}{0.000055}
\newcommand{\puckNFPredLossMean}{0.000072}
\newcommand{\puckNFPredLossStd}{0.000034}
\newcommand{\puckNFPredLossIQRLow}{0.000043}
\newcommand{\puckNFPredLossIQRHigh}{0.000103}
\newcommand{\puckNFPredLossMedianSci}{5.52 \times 10^{-5}}
\newcommand{\puckVAEPredLossMedian}{0.000092}
\newcommand{\puckVAEPredLossMean}{0.000099}
\newcommand{\puckVAEPredLossStd}{0.000036}
\newcommand{\puckVAEPredLossIQRLow}{0.000072}
\newcommand{\puckVAEPredLossIQRHigh}{0.000116}
\newcommand{\puckVAEPredLossMedianSci}{9.24 \times 10^{-5}}
\newcommand{\puckTestU}{29}
\newcommand{\puckTestP}{0.1212}
\newcommand{\puckTestPFormatted}{p = 0.12}

% =============================================================================
% DPMC Controller Task
% =============================================================================
\newcommand{\dpmcNFCatchMedian}{81.5\%}
\newcommand{\dpmcNFCatchMedianRaw}{0.815}
\newcommand{\dpmcNFCatchIQRLow}{79.2\%}
\newcommand{\dpmcNFCatchIQRHigh}{84.7\%}
\newcommand{\dpmcNFCatchMean}{74.6\%}
\newcommand{\dpmcNFCatchStd}{20.9\%}
\newcommand{\dpmcVAECatchMedian}{46.0\%}
\newcommand{\dpmcVAECatchMedianRaw}{0.460}
\newcommand{\dpmcVAECatchIQRLow}{33.2\%}
\newcommand{\dpmcVAECatchIQRHigh}{57.8\%}
\newcommand{\dpmcVAECatchMean}{46.2\%}
\newcommand{\dpmcVAECatchStd}{14.1\%}
\newcommand{\dpmcPhysicsCatchMedian}{89.0\%}
\newcommand{\dpmcPhysicsCatchMedianRaw}{0.890}
\newcommand{\dpmcPhysicsCatchIQRLow}{80.2\%}
\newcommand{\dpmcPhysicsCatchIQRHigh}{99.0\%}
\newcommand{\dpmcPhysicsCatchMean}{83.9\%}
\newcommand{\dpmcPhysicsCatchStd}{22.2\%}
\newcommand{\dpmcNFvsVAETestU}{90}
\newcommand{\dpmcNFvsVAETestP}{0.0028}
\newcommand{\dpmcNFvsVAETestPFormatted}{p = 0.003}
\newcommand{\dpmcNFvsPhysicsTestU}{25}
\newcommand{\dpmcNFvsPhysicsTestP}{0.0634}
\newcommand{\dpmcNFvsPhysicsTestPFormatted}{p = 0.06}

% =============================================================================
% Derived Metrics
% =============================================================================
\newcommand{\dpmcNFTransferEfficiency}{91.6\%}
\newcommand{\dpmcVAETransferEfficiency}{51.7\%}

% =============================================================================
% Body Schema Selectivity (Experiment 4)
% =============================================================================
\newcommand{\bodySchemaSelectivityMedian}{1.24}
\newcommand{\bodySchemaSelectivityMean}{1.20}
\newcommand{\bodySchemaSelectivityStd}{0.36}
\newcommand{\bodySchemaSelectivityIQRLow}{0.92}
\newcommand{\bodySchemaSelectivityIQRHigh}{1.41}
\newcommand{\bodySchemaNumSeeds}{10}
\newcommand{\bodySchemaTestW}{42.0}
\newcommand{\bodySchemaTestP}{0.0801}
\newcommand{\bodySchemaTestPFormatted}{p = 0.08}
\newcommand{\bodySchemaShoulderCSelectivityMedian}{1.12}
\newcommand{\bodySchemaShoulderCSelectivityMean}{1.17}
\newcommand{\bodySchemaShoulderCSelectivityStd}{0.40}
\newcommand{\bodySchemaShoulderRSelectivityMedian}{2.18}
\newcommand{\bodySchemaShoulderRSelectivityMean}{2.29}
\newcommand{\bodySchemaShoulderRSelectivityStd}{0.91}
\newcommand{\bodySchemaElbowCSelectivityMedian}{0.74}
\newcommand{\bodySchemaElbowCSelectivityMean}{0.83}
\newcommand{\bodySchemaElbowCSelectivityStd}{0.29}
\newcommand{\bodySchemaElbowRSelectivityMedian}{1.50}
\newcommand{\bodySchemaElbowRSelectivityMean}{1.74}
\newcommand{\bodySchemaElbowRSelectivityStd}{0.79}
\newcommand{\bodySchemaShoulderCTestP}{0.1611}
\newcommand{\bodySchemaShoulderCTestPFormatted}{p = 0.16}
\newcommand{\bodySchemaShoulderRTestP}{0.0020}
\newcommand{\bodySchemaShoulderRTestPFormatted}{p = 0.002}
\newcommand{\bodySchemaElbowCTestP}{0.9580}
\newcommand{\bodySchemaElbowCTestPFormatted}{p = 0.96}
\newcommand{\bodySchemaElbowRTestP}{0.0020}
\newcommand{\bodySchemaElbowRTestPFormatted}{p = 0.002}
\newcommand{\bodySchemaLossCorrelationRho}{-0.25}
\newcommand{\bodySchemaLossCorrelationP}{0.4888}
\newcommand{\bodySchemaLossCorrelationPFormatted}{p = 0.49}

% =============================================================================
% Failure Mode Analysis (Prediction Discontinuity)
% =============================================================================
\newcommand{\failureNFMeanDisp}{0.76}
\newcommand{\failureNFMedianDisp}{0.77}
\newcommand{\failureNFMaxDisp}{2.06}
\newcommand{\failureNFPNinetyFiveDisp}{1.25}
\newcommand{\failureNFTeleportRate}{0.0\%}
\newcommand{\failureNFSeqTeleportRate}{0.0\%}
\newcommand{\failureVAEMeanDisp}{0.84}
\newcommand{\failureVAEMedianDisp}{0.68}
\newcommand{\failureVAEMaxDisp}{21.97}
\newcommand{\failureVAEPNinetyFiveDisp}{1.68}
\newcommand{\failureVAETeleportRate}{1.9\%}
\newcommand{\failureVAESeqTeleportRate}{15.4\%}
\newcommand{\failureMaxDispRatio}{10.7}
\newcommand{\failureThreshold}{3.0}
\newcommand{\failureTestU}{23038}
\newcommand{\failureTestP}{0.0000}
\newcommand{\failureTestPFormatted}{p < 0.001}

% =============================================================================
% Sample Sizes
% =============================================================================
\newcommand{\numSeeds}{10}
\newcommand{\numEpisodesPerSeed}{100}

%% file: sections/introduction_cogsci_edited.tex
\section{Introduction}

A child learning to catch must solve two physics problems at once: where will
the ball land, and where will my hand be when I reach?
With practice, both predictions become automatic.

How does the brain pull this off?
Battaglia, Hamrick, and Tenenbaum proposed that humans possess an ``intuitive
physics engine'' (IPE) that runs mental simulations to predict physical outcomes
\citep{battaglia2013simulation}.
The framework accounts for human performance across diverse physical reasoning tasks \citep{hamrick2016inferring, ullman2017mind}.

Several computational frameworks have attempted to model physical intuition.
Object-based approaches like Interaction Networks \citep{battaglia2016interaction}
and the Neural Physics Engine \citep{chang2016compositional} learn to predict
physical dynamics over explicit object graphs.
These models capture important structure but operate on symbolic object
representations rather than raw visual input, and they do not integrate motor
commands.
To understand how the brain predicts physics while acting in the world, we need
models that integrate action.
\citet{grush2004emulation} proposed that the brain constructs internal emulators: forward models that predict the sensory consequences of motor commands and can run offline during imagery.
Such emulators could support task improvement away from the environment, as in mental practice \citep{jeannerod1995mental}.
World models from machine learning, which learn these predictions for planning and control, can be seen as computational implementations of this idea \citep{ha2018world}.
But these architectures omit cortical constraints.
Visual input is compressed through an encoder into a latent vector with no
spatial structure, a recurrent model predicts latent state transitions, and
finally a decoder reconstructs predictions.
In such architectures, a single weight matrix can connect any latent dimension
to any other, allowing information to jump discontinuously rather than propagate
locally.
In the physical world, a ball moves continuously through space, forces act on
adjacent objects, and trajectories unfold through spatial neighborhoods.
Latent-space world models have no such constraints.
Predicted objects can ``teleport'' across representational space from one timestep to the next.

In trajectory-prediction tasks, motion-sensitive area MT activates even without
visual motion \citep{ahuja2022role, ahuja2024monkeys}.
Neural patterns during prediction resemble those during actual motion
perception, as if physics simulation unfolds \textit{within} spatially organized
visual representations rather than in a separate abstract reasoning system.

We propose that the key architectural principle is \textbf{isomorphism}:
world models that preserve the spatial structure of sensory input.
The defining constraint is \textbf{locality}: nearby points in the world map to
nearby points in the representation, and information propagates through spatial
neighbors rather than ``teleporting'' via global connections.
When this constraint is enforced, physics prediction becomes a geometric
problem: a ball's future is a path through representational space, not an
abstract vector transition.

To test this hypothesis, we use neural fields, a simple
implementation of the isomorphic principle.
Introduced by Amari \citep{amari1977dynamics}, neural fields are spatially
organized recurrent networks where activity evolves through local lateral
interactions: each location influences only its spatial neighbors through a
connectivity kernel.
Neural fields have been used as models of cortical dynamics, capturing
phenomena from visual attention to working memory to motor planning.
Yet they have not been explored as world models for planning and control.
Neural fields are not the only possible isomorphic world model, and 2D tasks do not exhaust physical reasoning; here they provide a minimal, interpretable test of whether preserving spatial topology changes prediction, control, and motor-contingent representation.

To enable action-conditional prediction, we introduce \textit{motor-gated
channels}: populations whose activity is multiplicatively modulated by motor
commands.
The mechanism implements gain modulation, a common computational principle
in cortex whereby one signal multiplicatively scales neural responses to another
\citep{salinas2000gain}.
In posterior parietal cortex, gain fields enable sensorimotor integration by
combining visual input with motor signals \citep{andersen1997multimodal}.
Motor-gated channels implement this operation.

A motor-gated neural field can learn a world model that supports three cognitively relevant capacities --- physical prediction, learning a new task from internal simulation, and body-linked representation --- while respecting three neurobiological constraints: spatial maps, local lateral dynamics, and gain-like motor modulation.

\begin{figure}[!t]
    \centering
    \includegraphics[width=\linewidth]{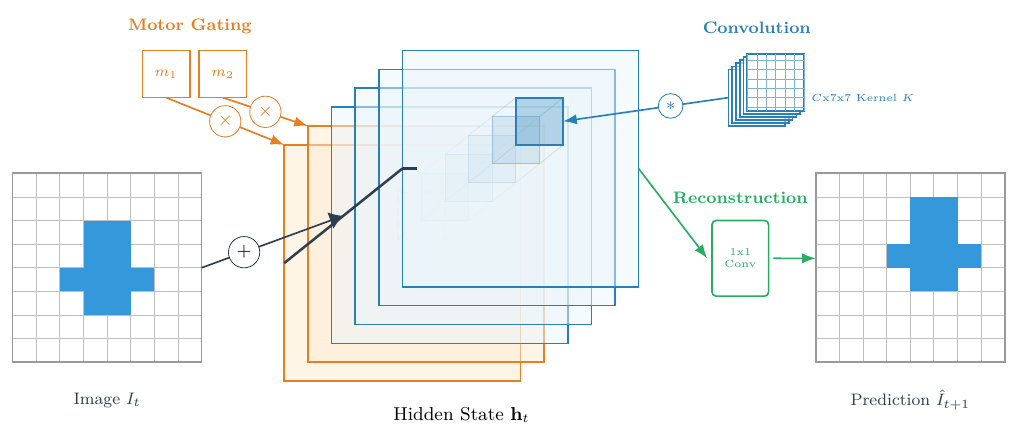}
    \caption{Neural field world model architecture. Visual input $I_t$ is added to the hidden state only during the first three timesteps of each sequence; thereafter, the field evolves autonomously through local lateral convolution ($K$) and recurrence. Motor commands $\mathbf{m}$ multiplicatively gate specific channels, implementing gain modulation. A 1x1 convolution reconstructs the visual prediction $\hat{I}_{t+1}$.}
    \label{fig:architecture}
\end{figure}

We provide preliminary evidence for this claim in three experiments:
\begin{enumerate}
    \item \textbf{Physics from Lateral Dynamics:} The constraint of local
    connectivity should make blind physical prediction proceed through
    continuous paths rather than discontinuous jumps (Experiment 1).
    \item \textbf{Transferable Motor Control:} A frozen world model should
    support offline learning of a new control task: gradients pass
    through long coherent rollouts, and the resulting policy transfers to real
    physics without policy training in the real environment (Experiment 2).
    \item \textbf{Body-Selective Encoding:} Motor-gated channels should develop
    arm-selective activity without explicit body labels,
    providing a minimal computational precursor to body schema
    \citep{gallagher2005body} (Experiment 3).
\end{enumerate}

%% file: sections/methods_cogsci_edited.tex
\section{Methods}

\subsection{Neural Field Architecture}

The neural field implements the locality constraint: information propagates through spatial neighbors rather than jumping arbitrarily across the representation.
At each timestep, the field state $\mathbf{h} \in \mathbb{R}^{C \times H \times W}$ updates through three influences: decay of current activity, lateral input from neighboring locations, and visual input from the environment.
A $7 \times 7$ convolutional kernel $K$ (randomly initialized, learned end-to-end with the rest of the world model) implements lateral connectivity: predicted motion must traverse intermediate positions just as objects traverse intermediate locations in physical space.
The model uses an Amari-style neural field update \citep{amari1977dynamics}:
\begin{equation}
    \mathbf{h}_{t+1} = \mathbf{h}_t + \frac{\Delta t}{\tau}\left(-\mathbf{h}_t + K * \text{ReLU}(\mathbf{h}_t) + W_{\text{in}} * \mathbf{I}_t\right)
\end{equation}
Visual predictions emerge through linear reconstruction: $\hat{\mathbf{I}}_t = W_{\text{out}} * \mathbf{h}_t$.

Each channel is a complete 2D activity map over the visual field.
To integrate motor commands, we designate the first $M$ maps as motor-gated, where $M$ matches the number of continuous motor commands ($M=4$ for the arm; the ballistic experiment uses no motor gating).
After each dynamics update, these channels are multiplicatively modulated by motor signals:
\begin{equation}
    \mathbf{h}_{t+1}^{(i)} = m_i \cdot \tilde{\mathbf{h}}_{t+1}^{(i)} \quad \text{for } i \in \{1, \ldots, M\}
\end{equation}
This implements gain modulation, the same computational principle by which posterior parietal cortex combines visual and motor signals \citep{salinas2000gain, andersen1997multimodal}.

\textbf{Neurobiological constraints.}
The cortical comparison is structural: we ask what follows when a world model is built from constraints common in sensory and sensorimotor cortex.
It preserves three such constraints: retinotopic maps maintain spatial correspondence with visual input \citep{wandell2007visual}; local lateral connectivity restricts interactions to spatial neighbors; and gain-like modulation couples motor commands to spatial activity.
ReLU and convolutional kernels are abstractions of threshold-linear activity and local connectivity.
Learning is treated separately from architecture: the simulations use backpropagation, while functionally similar credit assignment may be possible in cortical circuits \citep{lillicrap2020backpropagation}.

\subsection{Task Environments}

\textbf{Ballistic trajectory (Experiment 1).}
A ball moves under gravity in a $32 \times 32$ visual field, with random initial position and velocity.
The model observes the first 3 frames, then predicts the remainder without visual input.
Architecture: 16 channels (no motor gating).

\textbf{Musculoskeletal arm (Experiments 2--3).}
A planar double-pendulum arm operates in a $120 \times 45$ visual field with four motor commands following the Equilibrium-Point Hypothesis \citep{feldman1986equilibrium}.
For each joint, co-contraction (C) activates opposing muscles together to increase stiffness; reciprocal (R) shifts the equilibrium angle to produce movement direction.
The four commands are shoulder-C, shoulder-R, elbow-C, and elbow-R.
These C/R variables are scalar motor commands supplied at each timestep, not labels for visual regions.
A ball falls from a random horizontal position for catching.
Architecture: 32 channels (4 motor-gated).

\subsection{Training and Evaluation}

World models are trained with mean squared prediction loss (Adam optimizer, 10,000 epochs).
Frozen-simulator policy training (Experiment 2) has three stages.
First, the arm world model is learned from random motor-babbling trajectories using next-frame prediction.
Second, a small CNN+LSTM catch predictor is trained as a differentiable surrogate for the binary catch event.
It receives visual reconstructions generated by the frozen world model while replaying real-physics rollouts; per-timestep caught/missed labels come from the corresponding real-physics trajectory.
Third, the world model and catch predictor are frozen, and only the policy weights are updated.
During each simulated rollout, the policy observes the reconstructed visual field, outputs motor commands, the frozen world model rolls forward, and the frozen catch predictor scores the predicted trajectory.
The policy maximizes predicted catch probability through gradients that pass through both fixed modules.
Thus policy learning is optimization through a learned differentiable simulator, not model-free reinforcement learning from real-environment rewards or a rollout sampler for a reward-based learner.
Because gradients pass through long simulated trajectories, rollout coherence directly shapes the learned policy.
The trained policy deploys to real physics without fine-tuning.
Catch rate measures successful interceptions over 100 episodes.

For body schema analysis (Experiment 3), we compute a selectivity index comparing motor-gated channel activity over arm versus ball regions, normalized by reconstruction activity:
\[ \text{Selectivity} = \frac{\text{Motor}_\text{arm} / \text{Motor}_\text{ball}}{\text{Recon}_\text{arm} / \text{Recon}_\text{ball}} \]
Values greater than 1 indicate body-selective representation.

\textbf{Baseline.}
We compare against a VAE-LSTM inspired by \citet{ha2018world} because it represents the architectural contrast of interest: a latent-vector world model with global recurrent dynamics, where an unstructured recurrent state can mix information from any location.
A convolutional VAE compresses observations to a 32-dimensional latent space, an LSTM predicts latent dynamics, and a decoder reconstructs predictions.
Following \citet{ha2018world}, training proceeds in two stages (VAE first for 10,000 epochs, then frozen VAE plus LSTM for 10,000 epochs) because gradients through the decoder destabilize joint training.
This is not a parameter-matched benchmark; parameter count is descriptive, while the contrast is latent-vector versus spatially structured world-model format.
The neural field requires no staged training and uses 17--67$\times$ fewer parameters (13K vs 850K for ballistic; 50K vs 1.7M for arm), largely because the VAE-LSTM requires dense layers connecting flattened features to latent space.

%% file: sections/results_cogsci_edited.tex
\section{Results}

The three experiments test consequences of preserving spatial topology.
Experiment 1 asks whether local field dynamics can predict ballistic motion; Experiment 2 asks whether a frozen world model can support offline learning of a new catching policy; Experiment 3 asks whether the same motor-gated arm model contains body-selective structure without body labels.

\subsection{Experiment 1: Physics from Lateral Dynamics}

\begin{figure}[t]
    \centering
    \includegraphics[width=\linewidth]{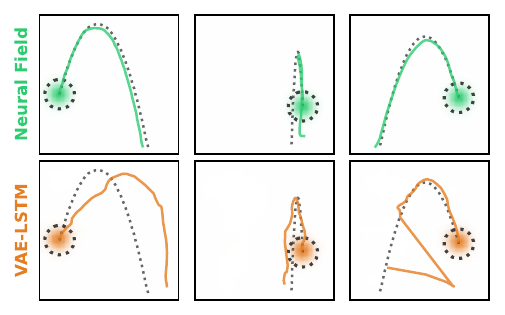}
    \caption{Trajectory prediction through internal dynamics. Each column overlays a roughly 50-timestep rollout in one plot, with successive predicted positions forming a path through time. Both models observe only the first three frames before blind prediction. The neural field (green) tracks smooth parabolic ground-truth arcs (gray dotted), while VAE-LSTM (orange) oscillates.}
    \label{fig:trajectory}
\end{figure}

Figure~\ref{fig:trajectory} shows ballistic prediction: after observing a ball for three frames, activity in the neural field continues along the parabolic arc even without visual input.
During observation, a localized bump tracks the ball; remove the input, and the bump keeps moving, following learned dynamics that approximate ballistic motion.

Across \numSeeds{} seeds, the neural field achieves median prediction loss of
$\ballisticNFPredLossMedianSci$ (IQR: [$\ballisticNFPredLossIQRLow$,
$\ballisticNFPredLossIQRHigh$]), compared to $\ballisticVAEPredLossMedianSci$
for VAE-LSTM (\ballisticTestPFormatted).
Beyond aggregate loss, VAE-LSTMs exhibit a failure mode that neural fields avoid.
We measured frame-to-frame displacement of predicted centroids during blind rollouts.
The neural field's maximum displacement was \failureNFMaxDisp{} pixels, with \failureNFSeqTeleportRate{} of sequences showing ``teleportation'' (jumps $>$\failureThreshold{} pixels).
VAE-LSTM showed \failureVAEMaxDisp{} pixels maximum (\failureMaxDispRatio$\times$ larger) and \failureVAESeqTeleportRate{} teleportation (\failureTestPFormatted).
Local connectivity bounds how far predictions can jump in a single timestep, a constraint absent from latent-space models.

\subsection{Experiment 2: Frozen-Simulator Policy Learning}

The harder test is offline task learning: can a frozen learned model
support training a catching policy that works in the real environment?
We freeze the trained neural field world model, then train a policy network to
catch falling balls using only the model's prediction of the visual field.
The policy receives the neural field reconstruction as input and outputs motor commands.
Training proceeds entirely offline within the world model: the model generates predicted
observations, the policy generates actions, and a fixed catch predictor (trained on real physics) scores each trajectory.
Gradients pass through the frozen world model to the policy, but do not update the world model itself.
This is not merely using imagined trajectories as samples; the frozen model is the differentiable simulator through which long-rollout errors shape the policy update.
No real-environment interaction or additional world-model training occurs during policy learning.

\begin{figure}[t]
    \centering
    \includegraphics[width=\columnwidth]{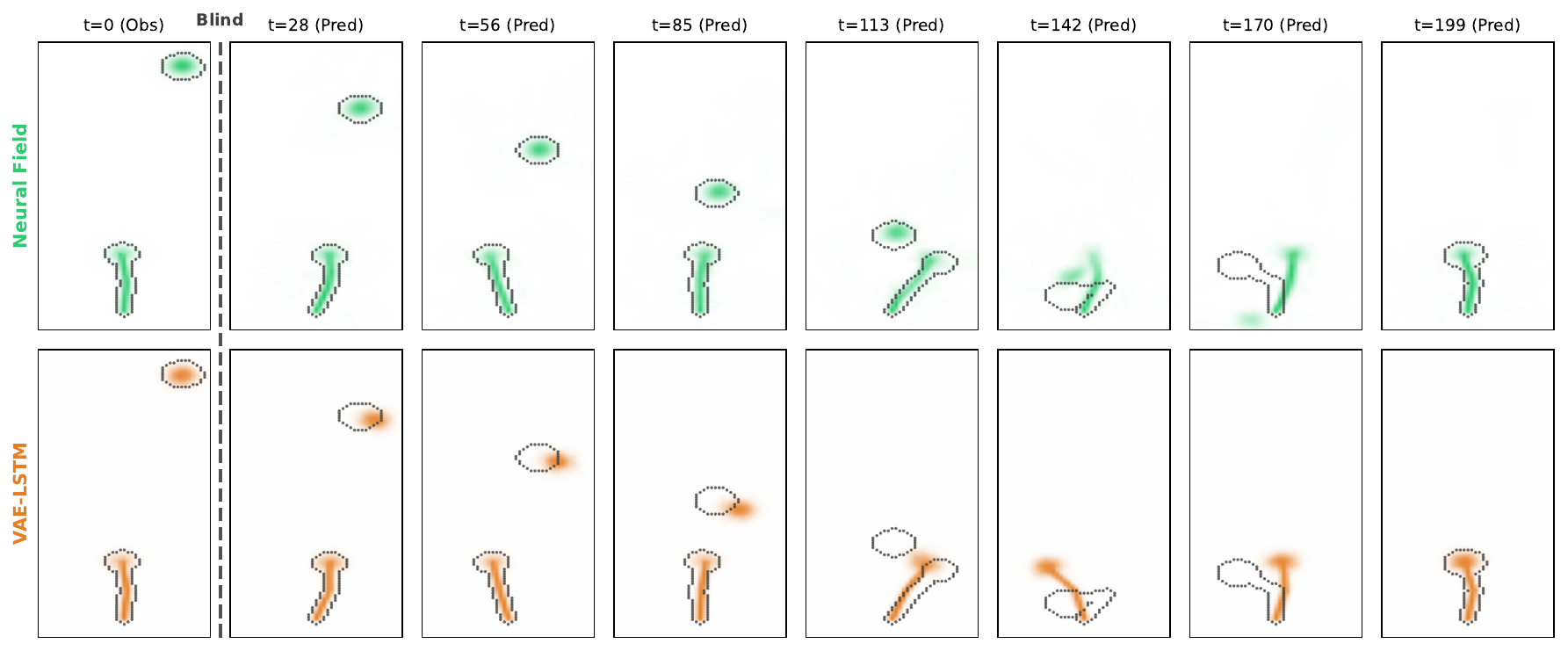}
    \caption{Visuomotor prediction in the arm catching task. Both models receive motor commands throughout, but visual input only during the first three frames (Obs). Ground truth shown as gray dots. The vertical dashed line marks the transition to blind prediction.}
    \label{fig:dpmc_trajectory}
\end{figure}

Figure~\ref{fig:dpmc_trajectory} shows the arm task: a double pendulum controlled by muscle-like actuators must catch a falling ball.
The neural field receives motor commands alongside visual input during observation, then predicts the visual consequences of continued motor activity during the blind phase.

\begin{figure}[t]
    \centering
    \includegraphics[width=\columnwidth]{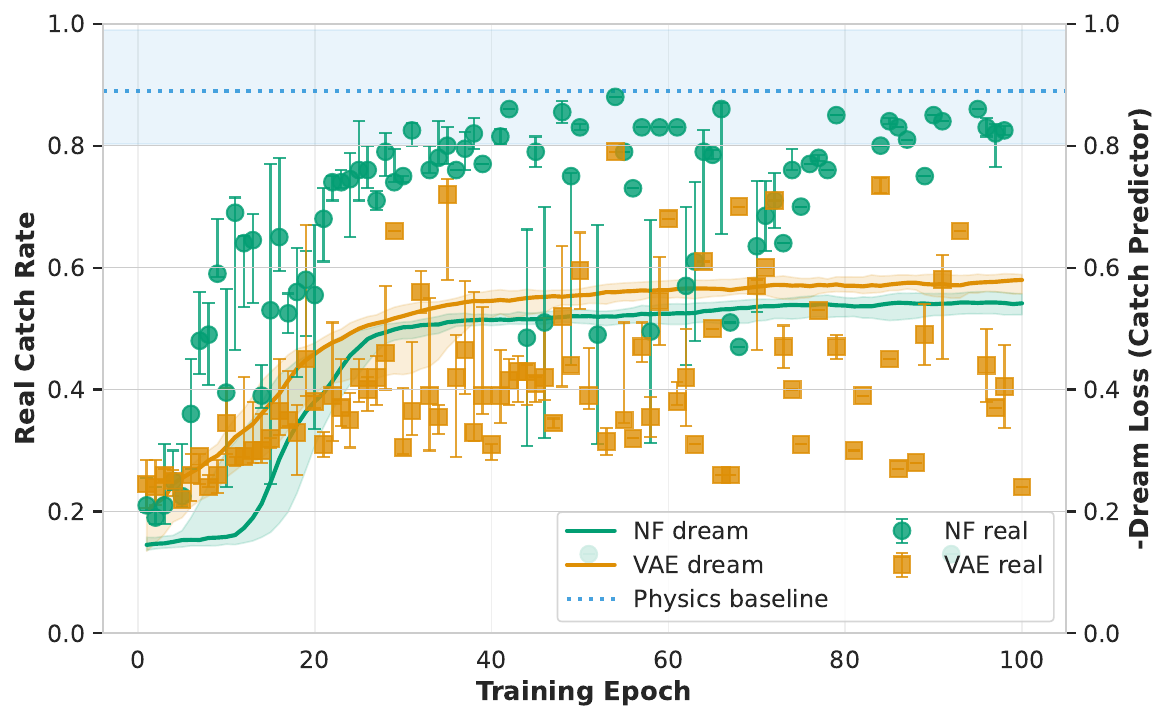}
    \caption{Frozen-simulator policy learning transfers to real physics. Lines show simulator-training catch-predictor value (right axis; negated loss, higher = better); this timestep-averaged training metric is capped near 60\% because a catch can only be registered after the ball reaches catching range, and is distinct from real catch rate. Points show real catch rate at evaluation (left axis). Neural field policies (green) achieve \dpmcNFCatchMedian{}, approaching the physics baseline (dotted). VAE-LSTM (orange) achieves comparable simulator-training performance but shows larger sim-to-real gap.}
    \label{fig:controller}
\end{figure}

Policies trained within the neural field achieve \dpmcNFCatchMedian{} catch
rate (IQR: [\dpmcNFCatchIQRLow, \dpmcNFCatchIQRHigh]) when deployed to the
\textit{real} physics environment, approaching the \dpmcPhysicsCatchMedian{}
achieved by policies trained directly on real physics
(Figure~\ref{fig:controller}).
VAE-LSTM policies achieve only \dpmcVAECatchMedian{}
(\dpmcNFvsVAETestPFormatted), roughly half the neural field's performance.

\begin{figure}[t]
    \centering
    \includegraphics[width=\linewidth]{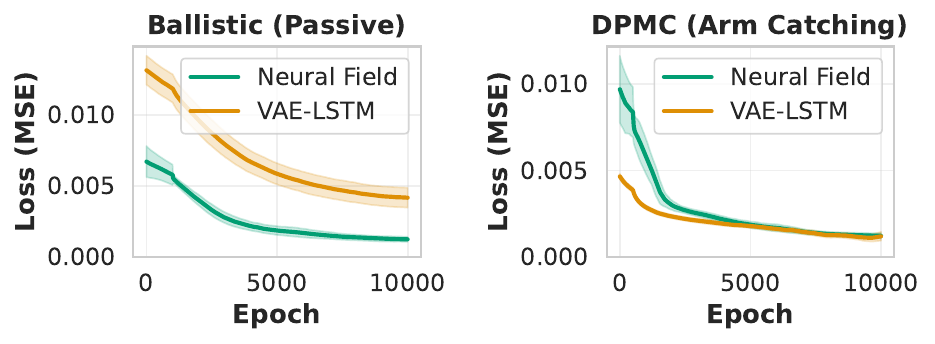}
    \caption{Training dynamics. Neural field (green) achieves lower final pixel-MSE loss than VAE-LSTM (orange) on the ballistic task and comparable loss on the arm task, despite using 17--67$\times$ fewer parameters. Task-relevant metrics are reported as centroid continuity and real catch rate. Shaded regions show IQR across seeds.}
    \label{fig:loss_curves}
\end{figure}

Prediction loss alone does not determine transfer quality: despite comparable
world model losses (Figure~\ref{fig:loss_curves}), the neural field supports substantially
better policy transfer.
MSE measures visual prediction fit; rollout continuity and real catch rate are the behaviorally relevant tests.

Why does spatial structure help?
The neural field's dynamics function as a spatial integrator: position and
velocity are represented in the location and movement of activity, and the
learned kernel implements discrete integration over time.
In latent-space models, these quantities must be disentangled from a compressed
representation.
The policy can therefore read spatial features directly from the field, rather than learning to decode them from a latent vector.

\subsection{Experiment 3: Emergent Body-Selective Encoding}

Body schema, the implicit representation of one's own body that enables
coordinated action, is studied as a distinct cognitive capacity
\citep{gallagher2005body}.
In the arm model, a computational precursor to body schema emerges as a
byproduct of visuomotor prediction.
The motor-gated architecture supplies action-modulated channels but does not specify which pixels represent the body or where selectivity should emerge; arm-versus-ball localization is measured only after training.
The finding provides computational support for developmental theories proposing
that infants discover their bodies through sensorimotor contingencies
\citep{bahrick1995intermodal, rochat1998self}.
Using the arm world model, we compare motor-gated channel activity over the arm
versus the ball, normalized by reconstruction activity.

\begin{figure}[t]
    \centering
    \includegraphics[width=\columnwidth]{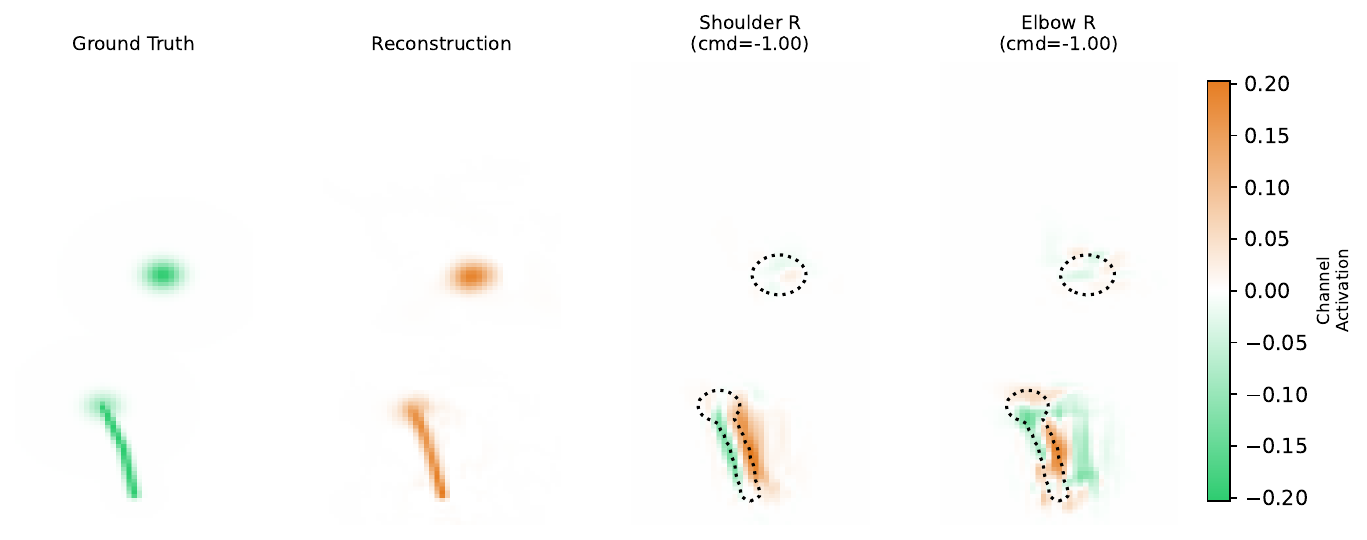}
    \caption{Emergent body-selective encoding in motor-gated channels. Reciprocal (R) channels show activation concentrated over the arm rather than the ball (dotted contours), despite no explicit training to distinguish body from world. Orange indicates positive activation, green negative.}
    \label{fig:body_schema}
\end{figure}

Reciprocal (R) motor channels, which control joint movement direction, show
significant arm selectivity (Figure~\ref{fig:body_schema}): shoulder R median
selectivity = \bodySchemaShoulderRSelectivityMedian{}
(\bodySchemaShoulderRTestPFormatted); elbow R median =
\bodySchemaElbowRSelectivityMedian{} (\bodySchemaElbowRTestPFormatted).
These channels are approximately $2\times$ more active over the arm than
reconstruction alone would require.
Co-contraction (C) channels show no significant selectivity (shoulder C:
\bodySchemaShoulderCTestPFormatted; elbow C: \bodySchemaElbowCTestPFormatted).

Why do R channels show selectivity while C channels do not?
Reciprocal commands produce large arm movements; to predict these, the motor
channel must represent where the arm is.
Co-contraction modulates stiffness with minimal immediate visual change.
The selectivity pattern reflects what each motor command \textit{does}.
Body selectivity emerges from the prediction objective, not explicit supervision.

%% file: sections/discussion_cogsci_edited.tex
\section{Discussion}

We proposed that motor-gated neural fields can learn world models that support three cognitively relevant capacities while respecting three neurobiological constraints: spatial maps, local lateral dynamics, and gain-like motor modulation.
Three results provide preliminary evidence.
First, local connectivity proves sufficient to learn ballistic physics: trajectories emerge from wave-like propagation through the field, and predictions traverse intermediate locations rather than teleporting as in latent-space models (Experiment 1).
Second, the spatial structure supports learning a new control task through a frozen differentiable simulator: policies trained through neural-field rollouts transfer to reality with a smaller sim-to-real gap than the latent baseline (Experiment 2).
Third, a computational precursor to body schema emerges from visuomotor prediction alone.
Motor-gated channels develop body-selective encoding without explicit body labels (Experiment 3).

\textbf{Relation to prior work.} The neural field offers a candidate implementation of
the ``intuitive physics engine'' proposed by Battaglia et al.\
\citep{battaglia2013simulation}: physics emerges from learned connectivity
rather than explicit rules.
Ahuja et al.\ found that area MT activates during trajectory prediction even
without visual motion \citep{ahuja2022role}.
Our neural field shows analogous behavior: during blind prediction, activity
moves through retinotopic space, providing a computational parallel to MT's
motion representation.
Multiplicative gating by motor signals implements gain modulation, the same computational principle found in posterior parietal cortex where motor signals modulate spatially organized sensory representations \citep{salinas2000gain, andersen1997multimodal}.
Neural fields have a rich history in cognitive science through Dynamic Field Theory (DFT), which hand-specifies Amari-type dynamics to model working memory, motor planning, and decision-making \citep{schoner2016dynamic}.
Our work departs from this tradition: rather than designing field dynamics for specific cognitive functions, we learn them end-to-end from sensorimotor prediction.
The neural field learns its own dynamics, including physics and body-selective encoding, through prediction error within this structured architecture.
On this view, computational principles identified by DFT can be learned when spatially structured networks predict the sensory consequences of action.

Neural fields and VAE-LSTMs can achieve similar loss,
but neural fields avoid the failure mode that affects VAE-LSTMs.
Local connectivity constrains errors to smooth spatial drift. Predictions cannot
jump arbitrarily across the neural field because information must propagate through spatial
neighbors.
Human multiple object tracking errors arise primarily from spatial proximity. For example, when targets approach within a few degrees, crowding causes confusions \citep{franconeri2010tracking}.
Neural fields with local connectivity should exhibit similar proximity-dependent interference, a prediction we leave for future work.

The VAE-LSTM comparison is a contrast between representational formats, not an exhaustive benchmark against all world models: spatially structured alternatives such as ConvLSTMs may share these benefits, which would support the broader isomorphism hypothesis.
Our claim is that preserving spatial topology changes how prediction errors unfold and whether a frozen model can train policies for new tasks through differentiable long rollouts.

\textbf{Implications.} Body-linked sensorimotor representation need not be explicitly labeled.
One might object that this is circular: gated channels will represent whatever moves with their gating signal.
The objection assumes, however, that the architecture supports stable visuomotor prediction in the first place; neither multiplicative gating nor purely local connectivity guarantee this.
The claim is therefore conditional: when motor-gated neural fields succeed at visuomotor prediction, body-selective encoding follows.

The result is one form of contingency detection \citep{bahrick1985detection}: given only action-conditional prediction, the distinction between controlled body and external object emerges from sensorimotor statistics.
Infants as young as three months preferentially attend to contingent visual
feedback from their own limb movements \citep{bahrick1995intermodal,
rochat1998self}. Our model offers a candidate mechanism by which this contingent attention could give rise to body-selective representations.
The motor-gated architecture is one such mechanism. Channels
gated by motor commands can develop representations of the body because the body
is what moves contingently with those commands.
This can be read as a simple comparator-like mechanism for agency \citep{blakemore2002abnormalities}, which proposes that the brain distinguishes self-caused from externally-caused sensations by comparing predicted sensory consequences against actual feedback.
Motor-gated channels generate action-conditional predictions; when these match observed motion, prediction error is low, while mismatch signals externally-caused motion.

Grush's emulation theory \citep{grush2004emulation} gives Experiment 2 its cognitive analogue: the brain constructs internal models (emulators) that can be run offline to simulate action outcomes.
During overt movement, these emulators run in parallel with actual sensorimotor loops; during imagery, they run alone.
Our neural field functions as such an emulator, a learned forward model that predicts sensory consequences of motor commands.
The key procedural distinction is that the emulator is not updated during policy learning, nor used merely to sample rollouts for a reward-based learner; it is the fixed differentiable system through which gradients train a new policy.
This imposes a stricter demand than short-term prediction: long rollouts must remain coherent because errors in the simulator's rollouts directly shape the policy gradients.
The training order also has a cognitive analogue: outcome recognition can be acquired from sparse exploratory feedback before a competent policy exists, then used to refine that policy offline.
Frozen-simulator learning succeeds because the emulator is accurate enough that policies trained offline transfer to reality.
Mental practice \citep{jeannerod1995mental} is a documented biological analogue, and dream-like rehearsal may be another: offline uses of internal forward models that improve action without engaging the body. We model this computational role, not sleep or imagery itself.

Latent-space and isomorphic models differ in how representation relates to world.
Latent-space models work as descriptions: experience is compressed into abstract codes, and spatial relations are recovered only through decoding.
Isomorphic models are constitutive rather than descriptive --- representation and world share the same geometric form.
To predict a trajectory is to propagate activity through space; to represent position is to activate a location.
The neural field is therefore directly interpretable: one can watch field dynamics unfold and read off the predicted position as activity in a particular location, not as a latent vector requiring decoding.
This aligns with enactivist accounts of perception
\citep{oregan2001sensorimotor}, where understanding arises from mastery of
sensorimotor contingencies rather than construction of internal pictures.
The motor-gated channels learn not a
symbol for ``arm'' but the sensorimotor contingency itself. They encode what changes when a particular action is commanded.
Because the representation is spatial, the answer is spatially readable, which may
explain why physical reasoning feels immediate rather than inferential.

\textbf{Limitations and future directions.}
Our demonstrations involve simple 2D environments. Object interaction, including occlusion, collision, and support, may require representing objects in separate layers with gated interactions, so they move independently while influencing each other through learned coupling.
More fundamentally, isomorphism does not need to mean pixel-isomorphism: the neural field could operate in a three-dimensional coordinate system that preserves the spatial structure of physical space rather than retinal space.
There is biological precedent: hippocampal place cells represent volumetric 3D space \citep{grieves2020place}, and parietal area V6A encodes reaching targets in depth coordinates \citep{hadjidimitrakis2014common}.
Visual input could project into this volumetric representation, and dynamics would unfold in a space where depth is explicit.
Such an architecture would remain isomorphic by preserving spatial adjacency in the physical world while handling the projective distortions that make 2D pixel space a poor model of 3D physics.

The observed benefits--stable policy gradients, successful simulator-to-real transfer, and body-selective encoding--likely arise from spatial adjacency plus action modulation rather than neural-field dynamics alone.
The broader class includes ConvLSTMs, cellular automata, and graph neural networks on spatial graphs--any architecture preserving spatial adjacency--and motor gating is one of several possible mechanisms for action integration.
Benchmarking this class to identify which architectural choices matter most for physics prediction is future work.

Intuitive physics supports but does not exhaust physical reasoning. The architecture we propose handles the fast, automatic simulation supporting perception and action; symbolic reasoning about physics likely requires additional mechanisms, and how neural field world models might interface with such systems remains an open question.

\textbf{Conclusion.} A motor-gated neural field treats physics, control, and body representation as one spatial problem: predicting what changes where when an agent acts.

%% file: references.bib
@article{amari1977dynamics,
  title={Dynamics of pattern formation in lateral-inhibition type neural fields},
  author={Amari, Shun-ichi},
  journal={Biological cybernetics},
  volume={27},
  number={2},
  pages={77--87},
  year={1977},
  publisher={Springer}
}

@book{schoner2016dynamic,
  title={Dynamic thinking: A primer on dynamic field theory},
  author={Sch{\"o}ner, Gregor and Spencer, John and DFT Research Group and others},
  year={2016},
  publisher={Oxford University Press}
}

@inproceedings{ha2018world,
  title={Recurrent World Models Facilitate Policy Evolution},
  author={Ha, David and Schmidhuber, J{\"u}rgen},
  booktitle={Advances in Neural Information Processing Systems},
  volume={31},
  pages={2451--2463},
  year={2018},
  publisher={Curran Associates, Inc.}
}

@article{grush2004emulation,
  title={The emulation theory of representation: Motor control, imagery, and perception},
  author={Grush, Rick},
  journal={Behavioral and Brain Sciences},
  volume={27},
  number={3},
  pages={377--396},
  year={2004},
  publisher={Cambridge University Press}
}

@article{oregan2001sensorimotor,
  title={A sensorimotor account of vision and visual consciousness},
  author={O'Regan, J Kevin and No{\"e}, Alva},
  journal={Behavioral and Brain Sciences},
  volume={24},
  number={5},
  pages={939--973},
  year={2001},
  publisher={Cambridge University Press}
}

@article{salinas2000gain,
  title={Gain modulation: A major computational principle of the central nervous system},
  author={Salinas, Emilio and Thier, Peter},
  journal={Neuron},
  volume={27},
  number={1},
  pages={15--21},
  year={2000},
  publisher={Elsevier}
}

@article{andersen1997multimodal,
  title={Multimodal representation of space in the posterior parietal cortex and its use in planning movements},
  author={Andersen, Richard A and Snyder, Lawrence H and Bradley, David C and Xing, Jing},
  journal={Annual Review of Neuroscience},
  volume={20},
  number={1},
  pages={303--330},
  year={1997},
  publisher={Annual Reviews}
}

@article{battaglia2013simulation,
  title={Simulation as an engine of physical scene understanding},
  author={Battaglia, Peter W and Hamrick, Jessica B and Tenenbaum, Joshua B},
  journal={Proceedings of the National Academy of Sciences},
  volume={110},
  number={45},
  pages={18327--18332},
  year={2013},
  publisher={National Acad Sciences}
}

@inproceedings{battaglia2016interaction,
  title={Interaction networks for learning about objects, relations and physics},
  author={Battaglia, Peter W and Pascanu, Razvan and Lai, Matthew and Rezende, Danilo and Kavukcuoglu, Koray},
  booktitle={Advances in Neural Information Processing Systems},
  volume={29},
  pages={4502--4510},
  year={2016}
}

@inproceedings{chang2016compositional,
  title={A compositional object-based approach to learning physical dynamics},
  author={Chang, Michael B and Ullman, Tomer and Torralba, Antonio and Tenenbaum, Joshua B},
  booktitle={International Conference on Learning Representations},
  year={2017}
}

@article{hamrick2016inferring,
  title={Inferring mass in complex scenes by mental simulation},
  author={Hamrick, Jessica B and Battaglia, Peter W and Griffiths, Thomas L and Tenenbaum, Joshua B},
  journal={Cognition},
  volume={157},
  pages={61--76},
  year={2016},
  publisher={Elsevier}
}

@article{ullman2017mind,
  title={Mind games: Game engines as an architecture for intuitive physics},
  author={Ullman, Tomer D and Spelke, Elizabeth and Battaglia, Peter and Tenenbaum, Joshua B},
  journal={Trends in Cognitive Sciences},
  volume={21},
  number={9},
  pages={649--665},
  year={2017},
  publisher={Elsevier}
}

@article{ahuja2022role,
  title={A role for visual areas in physics simulations},
  author={Ahuja, Aarit and Desrochers, Theresa M and Sheinberg, David L},
  journal={Cognitive Neuropsychology},
  volume={38},
  number={7-8},
  pages={425--439},
  year={2022},
  publisher={Taylor \& Francis},
  doi={10.1080/02643294.2022.2034609}
}

@article{ahuja2024monkeys,
  title={Monkeys engage in visual simulation to solve complex problems},
  author={Ahuja, Aarit and Rodriguez, Nadira Yusif and Ashok, Alekh Karkada and Serre, Thomas and Desrochers, Theresa M and Sheinberg, David L},
  journal={Current Biology},
  volume={34},
  number={24},
  pages={5635--5645},
  year={2024},
  publisher={Elsevier},
  doi={10.1016/j.cub.2024.10.026}
}

@article{blakemore2002abnormalities,
  title={Abnormalities in the awareness of action},
  author={Blakemore, Sarah-Jayne and Wolpert, Daniel M and Frith, Christopher D},
  journal={Trends in cognitive sciences},
  volume={6},
  number={6},
  pages={237--242},
  year={2002},
  publisher={Elsevier}
}

@article{jeannerod1995mental,
  title={Mental imagery in the motor context},
  author={Jeannerod, Marc},
  journal={Neuropsychologia},
  volume={33},
  number={11},
  pages={1419--1432},
  year={1995},
  publisher={Elsevier}
}

@article{lillicrap2020backpropagation,
  title={Backpropagation and the brain},
  author={Lillicrap, Timothy P and Santoro, Adam and Marris, Luke and Akerman, Colin J and Hinton, Geoffrey},
  journal={Nature Reviews Neuroscience},
  volume={21},
  number={6},
  pages={335--346},
  year={2020},
  publisher={Nature Publishing Group}
}

@article{franconeri2010tracking,
  title={Tracking multiple objects is limited only by object spacing, not by speed, time, or capacity},
  author={Franconeri, Steven L and Jonathan, Stephen V and Scimeca, Jason M},
  journal={Psychological Science},
  volume={21},
  number={7},
  pages={920--925},
  year={2010},
  publisher={Sage Publications}
}

@article{wandell2007visual,
  title={Visual field maps in human cortex},
  author={Wandell, Brian A and Dumoulin, Serge O and Brewer, Alyssa A},
  journal={Neuron},
  volume={56},
  number={2},
  pages={366--383},
  year={2007},
  publisher={Elsevier}
}

@incollection{bahrick1995intermodal,
  title={Intermodal origins of self-perception},
  author={Bahrick, Lorraine E},
  booktitle={The self in infancy: Theory and research},
  editor={Rochat, Philippe},
  series={Advances in Psychology},
  volume={112},
  pages={349--373},
  year={1995},
  publisher={North-Holland/Elsevier}
}

@article{rochat1998self,
  title={Self-perception and action in infancy},
  author={Rochat, Philippe},
  journal={Experimental Brain Research},
  volume={123},
  number={1},
  pages={102--109},
  year={1998},
  publisher={Springer}
}

@article{bahrick1985detection,
  title={Detection of intermodal proprioceptive--visual contingency as a potential basis of self-perception in infancy},
  author={Bahrick, Lorraine E and Watson, John S},
  journal={Developmental Psychology},
  volume={21},
  number={6},
  pages={963--973},
  year={1985},
  publisher={American Psychological Association}
}

@book{gallagher2005body,
  title={How the Body Shapes the Mind},
  author={Gallagher, Shaun},
  year={2005},
  publisher={Oxford University Press}
}

@article{feldman1986equilibrium,
  title={Once more on the equilibrium-point hypothesis ($\lambda$ model) for motor control},
  author={Feldman, Anatol G},
  journal={Journal of Motor Behavior},
  volume={18},
  number={1},
  pages={17--54},
  year={1986},
  publisher={Taylor \& Francis}
}

@article{grieves2020place,
  title={The place-cell representation of volumetric space in rats},
  author={Grieves, Roddy M and Jedidi-Ayoub, Selim and Mishchanchuk, Karyna and Liu, Anyi and Renaudineau, Sophie and Jeffery, Kate J},
  journal={Nature Communications},
  volume={11},
  number={1},
  pages={789},
  year={2020},
  doi={10.1038/s41467-020-14611-7},
  publisher={Nature Publishing Group}
}

@article{hadjidimitrakis2014common,
  title={Common neural substrate for processing depth and direction signals for reaching in the monkey medial posterior parietal cortex},
  author={Hadjidimitrakis, Kostas and Berber, M Farzaneh and Sheth, Archana and Bhansali, Priya and Breveglieri, Rossella and Bosco, Annalisa and Fattori, Patrizia and Galletti, Claudio},
  journal={Cerebral Cortex},
  volume={24},
  number={6},
  pages={1645--1657},
  year={2014},
  publisher={Oxford University Press}
}
